\documentclass[letterpaper, 10 pt, conference]{ieeeconf}  

\IEEEoverridecommandlockouts                              
                                                          
\usepackage{cite}
\usepackage{amsmath,amssymb,amsfonts}
\usepackage{algorithmic}
\usepackage{graphicx}
\usepackage{textcomp}
\usepackage{xcolor}
\def\BibTeX{{\rm B\kern-.05em{\sc i\kern-.025em b}\kern-.08em
    T\kern-.1667em\lower.7ex\hbox{E}\kern-.125emX}}

\usepackage{tikz}
\usepackage{textcomp}
\usepackage{hyperref}
\usepackage{lipsum}

\usepackage[acronym]{glossaries}
\usepackage{glossaries-prefix}
\usepackage{url}
\usepackage{tabularx}
\usepackage{multirow}
\usepackage{lipsum}
\usepackage{colortbl}
\usepackage[figuresright]{rotating}
\usepackage{lscape}
\usepackage{graphicx}
\usepackage{subfig}
\usepackage{caption}
\usepackage{adjustbox}
\usepackage{rotating}
\usepackage{balance}
\usepackage{booktabs}

\newcommand{\cready}[1]{#1}

\definecolor{red3}{RGB}{136,0,3}
\definecolor{comment}{named}{red3}
\definecolor{lightgrey}{RGB}{230,230,230}

\newacronym[
    prefixfirst={an\ },
    prefix={an\ }
    ]{ADS}{ADS}{Automated Driving System}
\newacronym[
    longplural={Advanced Driver Assistance System},
    shortplural={ADASs},
    prefixfirst={an\ },
    prefix={an\ }
    ]{ADAS}{ADAS}{Advanced Driver Assistance System}
\newacronym[
    longplural={Advanced Driver Assistance Systems / Automated Driving},
    shortplural={ADASs/AD},
    prefixfirst={an\ },
    prefix={an\ }
    ]
    {ADAS/AD}{ADAS/AD}{Advanced Driver Assistance System / Automated Driving}
\newacronym[
    prefixfirst={an\ },
    prefix={an\ }
    ]{AEB}{AEB}{Autonomous Emergency Braking} 
\newacronym{CAD}{CAD}{Computer Aided Design}
\newacronym{HAV}{HAV}{Highly Automated Vehicle}
\newacronym{IEEE}{IEEE}{Institute of Electrical and Electronics Engineers}
\newacronym{ITSC}{ITSC}{International Conference on Intelligent Transportation Systems}
\newacronym{IV}{IV}{Intelligent Vehicle Symposium}
\newacronym{LKAS}{LKAS}{Lane-Keeping Assistant System}
\newacronym[
    prefixfirst={a\ },
    prefix={an\ }
    ]{ODD}{ODD}{Operational Design Domain}
\newacronym{OEM}{OEM}{Original Equipment Manufacturer}
\newacronym{SBT}{SBT}{Scenario-based Testing}
\newacronym{SOTIF}{SOTIF}{"Safety of the Intended Functionality"}
\newacronym[
    prefixfirst={a\ },
    prefix={an\ }
    ]
    {SUT}{SUT}{System Under Test}
\newacronym[
    prefixfirst={a\ },
    prefix={an\ }
    ]
    {SLR}{SLR}{Systematic Literature Review}
\newacronym{VaV}{V\&V}{Verification and Validation}

\newcommand\copyrighttext{%
      \footnotesize \textcopyright 2023 IEEE. Personal use of this material is permitted.
      Permission from IEEE must be obtained for all other uses, in any current or future
      media, including reprinting/republishing this material for advertising or promotional
      purposes, creating new collective works, for resale or redistribution to servers or
      lists, or reuse of any copyrighted component of this work in other works.
      }
      
\newcommand{\copyrightnotice}{%
    \begin{tikzpicture}[remember picture, overlay]
        \node[anchor=south,yshift=11pt] at (current page.south) {\fbox{\parbox{\dimexpr\textwidth-\fboxsep-\fboxrule\relax}{\copyrighttext}}};
    \end{tikzpicture}%
}

\begin{document}

\title{\LARGE \bf
    Is Scenario Generation Ready for SOTIF? \\ A Systematic Literature Review
}


\author{Lukas Birkemeyer$^{1}$, Christian King$^{2}$, and Ina Schaefer$^{3}$
\thanks{$^{1}$Lukas Birkemeyer is with Technical University Braunschweig, Braunschweig, Germany
        {\tt\small l.birkemeyer@tu-braunschweig.de}}%
\thanks{$^{2}$Christian King is with FZI Forschungszentrum Informatik, Karlsruhe, Germany
        {\tt\small king@fzi.de}}%
\thanks{$^{3}$Ina Schaefer is with Karlsruhe Institute of Technology, Karlsruhe, Germany
        {\tt\small ina.schaefer@kit.edu}}%
}

\maketitle

\begin{abstract}
    Scenario-based testing is considered state-of-the-art to verify and validate Advanced Driver Assistance Systems or Automated Driving Systems. Due to the official launch of the SOTIF-standard (ISO 21448), scenario-based testing becomes more and more relevant for releasing those Highly Automated Driving Systems. However, an essential missing detail prevent the practical application of the SOTIF-standard: How to practically generate scenarios for scenario-based testing? In this paper, we perform a Systematic Literature Review to identify techniques that generate scenarios complying with requirements of the SOTIF-standard. We classify existing scenario generation techniques and evaluate the characteristics of generated scenarios wrt. SOTIF requirements. We investigate which details of the real-world are covered by generated scenarios, whether scenarios are specific for a system under test or generic, and whether scenarios are designed to minimize the set of unknown and hazardous scenarios. We conclude that scenarios generated with existing techniques do not comply with requirements implied by the SOTIF-standard; hence, we propose directions for future research.
\end{abstract}



\section{Introduction}
    \copyrightnotice
    In recent years, the automation of \gls{ADAS} increases aiming for more safety and comfort in road traffic. The gap between highly automated \gls{ADAS} and \gls{ADS} seems to disappear. 
    Established methods as provided in ISO~26262 \cite{iso26262} are no longer sufficient for the \gls{VaV}. The standard \gls{SOTIF} (ISO~21448) \cite{iso21448} and the UN/ECE regulation R157 \cite{uneceR157} require \gls{SBT}.
    
    \gls{SBT} investigates the behavior of a \gls{SUT} in a driving scenario. \cready{The} scenario consists of a sequence of static scenes which are \textit{"snapshot[s] of the environment including the scenery [and] dynamic elements, [...]
    and the relationships among those entities."} \cite{iso21448, ulbrich2015defining}. The scenery describes lane information, static elements (e.g., traffic lights) and weather conditions. Dynamic elements are road users such as pedestrians or other vehicles.
    There are different abstraction levels of scenarios \cite{Bagschik2018a}: \textit{Logical scenarios} describe the main semantics of a scenario; logical scenarios contain ranges of possible parameters -- e.g., a vehicle drives in a city with [20;60]~km/h and a pedestrian crosses the street with [2;6]~km/h.
    To perform \gls{SBT}, \cready{logical scenarios need to be concretized} into a lower abstraction level (\textit{concrete scenario}) by defining concrete parameter values~\cite{annexC(2022)5402}.
    In this paper, we use the term \textit{scenario generation} to describe the process of deriving concrete scenarios from logical scenarios.

    The \gls{SOTIF}-standard was officially published in July 2022; a preliminary version (ISO/PAS~21448)~\cite{iso21448pas} is public since January 2019. Current literature suggesting scenario generation approaches is often motivated by \gls{SOTIF} but does not discuss the suitability of generated scenarios wrt. the standard.
    For example, \gls{SOTIF} requires that the operational domain and foreseeable misuse are covered in the \gls{SBT} process. \gls{SOTIF} also requires to focus on scenarios that lead to hazardous behavior of the \gls{SUT}. Still, \gls{SOTIF} misses details hindering a proper release of highly automated \gls{ADAS} and \gls{ADS} \cite{birkemeyer2022fahren}.
    \gls{SOTIF} neither specifies techniques to generate scenarios, nor metrics to assess whether generated logical or concrete scenarios (individual or as scenario suite) fulfill the \gls{SOTIF} requirements. Hence, for testers it is not clear how to design \textit{good} or \textit{better} scenarios for testing; for manufacturers, it is not clear when to release \gls{ADS} with acceptable residual risk; for legislators, it is unclear which scenario suites are suitable to allow \gls{ADS} on public streets. Without solving open challenges in scenario generation processes, a proper release of \gls{ADS} according to the \gls{SOTIF}-standard is not possible. 

    Current research activities \cite{gambi2019generating, montanari2021maneuver,hauer2020clustering, krajewski2019beziervae, nalic2019development, de2017assessment, althoff2018automatic, birkemeyer2022feature, rocklage2017automated, paranjape2020modular, yang2020multi, wang2020behavioral, medrano2019abstract} present and evaluate approaches to generate scenarios. However, they do not evaluate whether generated scenarios comply with \gls{SOTIF} requirements. In this paper, we analyze characteristics of scenario generation approaches and pursue the research goal:
    Identification of approaches that generate scenarios complying with requirements of \gls{SOTIF} to practically apply scenario-based testing.
    
    In this paper, we perform \pgls{SLR} and make the following contributions: 
    \begin{itemize}
        \item We structure existing scenario generation approaches according to their scenario generation technique.
        \item We analyze which existing scenario generation techniques generate scenarios that comply with SOTIF requirements.
        \item We propose directions for future work on scenario generation complying with \gls{SOTIF} requirements.
    \end{itemize}


\section{Background}

    In this section, we introduce the \gls{SOTIF}-standard~\cite{iso21448} to verify and validate \gls{ADAS}/\gls{ADS}. We also provide fundamental knowledge of a scenario structure that we use in this \gls{SLR} \cready{to make the paper self-contained}.

    \subsection{SOTIF}
    
        The \acrfull{SOTIF}-standard (ISO~21448~\cite{iso21448}) complements the established ISO~26262~\cite{iso26262}.
        The \gls{SOTIF}-standard proposes \gls{SBT} to avoid unintended behavior of \pgls{ADAS}/\gls{ADS} that is traceable to the interaction of the driving function with the environment.
        \gls{SOTIF} classifies scenarios into four classes, deciding whether a scenario is \textit{known} or \textit{unknown} and \textit{hazardous} or \textit{not hazardous}. "Hazardous scenarios are scenarios causing \textit{hazardous behavior}" \cite{iso21448}. Hazardous behavior, in turn, potentially leads to \textit{harm}. The goal of \gls{SOTIF} is to minimize the set of \textit{unknown} and \textit{hazardous} scenarios. \gls{SOTIF} processes analyze \glspl{SUT} for hazards and functional insufficiencies as well as related triggering conditions to define a \acrfull{VaV} strategy. 
        Based on the \gls{SBT} results, \gls{SOTIF} estimates whether the "\textit{residual risk}" or the "\textit{likelihood of encountering an unknown scenario leading to hazardous behaviour [is] sufficiently small}" \cite{iso21448}.
        Methods or metrics to determine the size of the hazardous scenario set (i.e. the number and occurrence probability of hazardous scenarios) are not provided; instead, \gls{SOTIF} suggests 16 methods to evaluate residual risk (cf. Table~11~\cite{iso21448}). However, knowledge regarding completeness wrt. the real-world is necessary for a proper safety argumentation. The \gls{SOTIF}-standard requires scenarios to fulfill two essential requirements:
        
        \textit{Req-1: Generated scenarios have to model the overall \acrshort{ODD} and foreseeable misuse scenarios.}
        According to \gls{SOTIF}, we have to consider \gls{ODD}-boundaries and foreseeable misuse in the \gls{VaV} process \cite{iso21448}.
        The \gls{SOTIF}-standard defines an \gls{ODD} as "\textit{specific conditions under which a given driving automation system is designed to function}"~\cite{iso21448}. These conditions include a precise definition of, e.g., the type of road, weather conditions, traffic events. To make sure, that a \gls{SUT} properly identifies whether it is within or outside of \pgls{ODD}, we need to test inside and outside scenarios \cite{annexC(2022)5402}. Wrt. \gls{ADAS} (and partly \gls{ADS}) a limited \gls{ODD} and foreseeable misuse might be sufficient. Regarding full automated driving, \gls{SOTIF} implicitly requires generated scenarios that cover the real-world. 

        \textit{Req-2: Generated scenarios contribute to minimizing the set of (unknown) hazardous scenarios.} The goal of \gls{SOTIF} activities is to determine the residual risk resulting from unknown scenarios~\cite{iso21448}. Minimizing the number of (unknown) hazardous scenarios leads to a lower risk.
    
    \subsection{Structure of a Scenario}

        To determine covering completeness of generated scenarios, we need to compare scenarios. Bagschik~et~al.~\cite{Bagschik2018a} structure, in accordance with \gls{SOTIF}, a scenario in five levels. The first three levels (\textit{E1}--\textit{E3}) define the static vehicle environment, while the levels \textit{E4} and \textit{E5} define dynamic elements and weather conditions. Scenario level \textit{E1} describes the trajectory and surface of a road. Scenario level \textit{E2} defines the equipment of the road such as lane marking, traffic lights and signs. The third level \textit{E3} contains temporal adjustments of the scenario levels \textit{E1} and \textit{E2}. These temporal adjustments last longer than one day, containing, e.g., construction sites or road closures. Scenario level \textit{E4} describes dynamic elements such as vehicles, pedestrians or animals and level \textit{E5} contains weather conditions such as snow or rain.


\section{State-of-the-Art \& Problem Statement}

    Current literature \cite{menzel2019functional, stellet2015testing, annexC(2022)5402, nalic2020scenario} suggests a binary classification of scenario generation techniques: (1) data-driven and (2) knowledge-based. Data-driven techniques derive scenarios from databases and knowledge-based techniques generate scenarios based on expert knowledge e.g., in the form of traffic regulations.
    This classification does not cover existing scenario generation approaches that use optimization techniques such as \cite{althoff2018automatic, gangopadhyay2019identification}.
    In these approaches, a fitness function is defined by expert knowledge, but the actual selection of concrete parameters is neither data-driven nor knowledge-based. 
    Ding et al. \cite{ding2023survey} consider a third scenario generation technique: adversarial scenario generation. This includes optimization techniques but does not cover combinational scenario generation as provided in \cite{birkemeyer2022feature, rocklage2017automated}. 
    Birkemeyer et al. \cite{birkemeyer2022feature} suggest, but do not validate a more fine granular classification of scenario generation techniques that differentiates between (1) random, (2) data-driven, (3) optimization, and (4) combinational scenario generation. In random scenario generation, parameters are randomly selected, while data-driven approaches derive scenarios from databases as in the binary classification above. Optimization techniques iteratively select and evaluate scenario parameters; combinational scenario generation divides scenarios into atomic blocks and combines them systematically. 
    
    Current surveys and \glspl{SLR} focus on \gls{SBT} to verify and validate \gls{ADAS}/\gls{ADS} \cite{ma2022verification, zhang2022finding, zhong2021survey, nascimento2019systematic, alawadhi2020systematic, jing2020agent, tahir2020coverage, karunakaran2022challenges, nalic2020scenario, schutt20231001}. These reviews partially focus on the scenario generation process; However, the scope is not directed to a classification that covers all existing techniques to generate concrete scenarios except \cite{ma2022verification} identifies a classification similar to \cite{birkemeyer2022feature} \cready{and Schütt et al. \cite{schutt20231001} propose a taxonomy of scenario acquisition categories and their relations.}
    \cready{T}o the best of our knowledge, there is no survey or \gls{SLR} discussing suitability of scenario generation approaches regarding \gls{SOTIF}.
    Thus, this \gls{SLR} aims to identify and quantitatively justify a fine granular classification of existing scenario generation techniques based on current literature. It also discusses whether existing scenario generation techniques are suitable to generate scenarios that meet requirements implied by \gls{SOTIF} which is novel.


\section{Planning the literature review} \label{sec:research}

\begin{figure*}
        \centering
        \includegraphics[width=\textwidth]{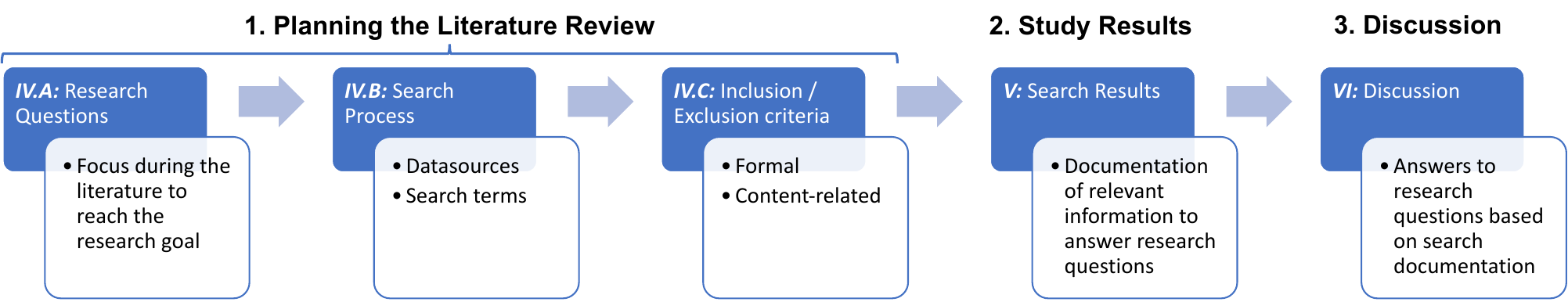}
        \caption{
        Guideline to perform a \acrfull{SLR} according to Kitchenham~\cite{kitchenham2004procedures}}
        \label{fig:SLR_process}
    \end{figure*}

    Our \gls{SLR} is structured according to guidelines provided by Kitchenham \cite{kitchenham2004procedures} (cf. Figure \ref{fig:SLR_process}). In this section, we plan the literature review by deriving research questions from our research goal (IV.A), specifying the search process (IV.B) and defining exclusion criteria (IV.C). The following section~\ref{sec:study_results} describes the execution of the study. Section~\ref{sec:discussion} discusses the findings regarding our research questions.

    \subsection{Research Questions} \label{sec:questions}

        Aiming for general findings, we investigate: \textbf{RQ1: Which techniques are used to generate scenarios for scenario-based testing?} The objective is to cluster existing approaches according to the scenario generation technique that is used.        
        
        To reach our research goal -- identification of approaches that generate scenarios to practically verify and validate \gls{ADAS}/\gls{ADS} according to \gls{SOTIF} -- we define research question \textbf{RQ~2:~Do existing approaches generate scenarios that comply with the requirements of SOTIF?}
        We subdivide RQ~2 into the following three aspects.
        
        Completeness in the sense of modeling the real-world is an infeasible objective to reach in \gls{SBT}~\cite{iso21448}. However, completeness of the generated scenarios is essential to calculate residual risks for a proper release of \gls{ADAS}/\gls{ADS}. Since we have to consider scenarios that are inside and outside of the \gls{SUT}'s \gls{ODD} to avoid foreseeable misuse (cf. Req-1), we are interested in the aspects that are considered by existing scenario generation approaches. We ask: \textbf{RQ~2.1:~Which details of the real-world are covered by generated scenarios?} The objective is to determine which scenario levels (according to Bagschik~et~al.~\cite{bagschik2018wissensbasierte}) are derived from existing scenario generation techniques. For approaches that address all scenario levels, we identify whether elements of a scenario level are overlooked to potentially point out incomplete coverage in modeling the real-world.
        Thus, the SOTIF scenario classification and, subsequently, minimizing the set of (unknown) hazardous scenarios (cf. \textit{Req-2}) significantly depends on the interaction of both, \gls{SUT} and scenario (including triggering condition). We examine both aspects independently in the research questions RQ~2.2 and RQ~2.3. First, we focus on the correlation of scenario generation and \gls{SUT}:
        \textbf{RQ~2.2:~Are generated scenarios \gls{SUT}-specific?} The objective is to determine whether system information is required to generate scenarios.
        Second, we focus on the scenario and potential triggering conditions. We are interested in:
        \textbf{RQ~2.3: Are scenarios designed to trigger hazardous behavior?}
        The objective is to classify generated scenarios according to the four SOTIF classes. We analyze, whether scenario generation techniques select scenarios that represent an overall scenario space or whether scenarios are explicitly designed to be hazardous. In contrast to RQ~2.1, we do not focus on \gls{ODD}-completeness.
            
    \subsection{Search process} \label{sec:search_process}
    
        We systematically identify articles that are relevant for scenario generation by defining a search string and collecting data from search engines by searching for the search string in abstracts (cf. Figure \ref{fig:SLR_process}, IV.B). We determine terms and synonyms that enclose the literature of interest.
        We truncate terms and use placeholders to cover grammatical use cases. We define the search string: \textit{(scenario* OR scenario-based) AND (generat* OR creat* OR select*) AND ("driver assistance" OR "self driving" OR "self-driving" OR (("autonomous" OR "automated") AND ("driving" OR "car" OR "vehicle"))) AND (verif* OR valid* OR test*)}

        We collect literature from search engines that are commonly used for reviews of scientific literature. In particular, we focus on \textit{ACM Digital Library}, \textit{IEEE Xplore} and \textit{Google Scholar}. In contrast to the other search engines, \textit{Google Scholar} is not limited to a single library. Moreover, it provides literature that matches the exact search string and related literature. This results in an enormous number of (potentially irrelevant) results. To handle this problem, we start with Google Scholars \textit{most relevant} results and scan abstracts and titles manually. We collect articles that deal with the automotive domain and testing of \gls{ADAS}/\gls{ADS} or present simulation tools to generate scenarios until 15 articles in a row do not match these criteria.

    \subsection{Exclusion criteria}
        To make sure, that we only consider relevant articles, we define inclusion/exclusion criteria (cf. Figure~\ref{fig:SLR_process}, IV.C).
        \cready{As a \textit{formal} criterion, we require that relevant articles are peer-reviewed contributions such as conference papers or journal articles ensuring that our study is based on comparable, high-quality articles. Thus, we explicitly exclude magazine articles and extended abstracts. We also require that articles are written in English and publicly available.}
        \textit{Content-related criteria:}
        We require that articles focus on scenario generation (i.e., the transformation from logical to concrete scenarios) to perform \gls{SBT} in the automotive context.
        Another content-related criterion is that a study uses the definition of a scenario according to Ulbrich~et~al.~\cite{ulbrich2015defining}, the \gls{SOTIF}-standard~\cite{iso21448} or similar. We explicitly include articles that present a simulation tool that supports scenario generation and simulation of scenarios, although they do not explicitly present a scenario generation process. This comes with the idea that simulation tools transform discrete input parameters into a continuous physical representation.


\section{Study Results} \label{sec:study_results}

     We consider articles that are published before 2023 and present, the search results by sharing the numbers of papers that we have found, included and excluded. We substantiate these results by conducting each step with the four-eye principle of two independent researchers.

    \textit{Search Results:}
        According to the search process, we collect a total of 892~articles. The search engine of the \textit{ACM Digital Library} provided 122~articles, the search engine of \textit{IEEE Xplore} identified 641~articles and \textit{Google Scholar} provided \cready{-- combined with manual identification --} 129~articles.

    \textit{Inclusion and Exclusion:}
        We collect the results from the search engines and remove duplicates resulting in a set of 779~articles. We formally exclude 21~articles since they do not comply with our formal criteria.
        Subsequently, we faithfully check the remaining 758~articles according to content-related criteria; maintaining 159~relevant articles.
        The excluded articles mainly focus on \cready{developing/improving} \gls{ADAS}/\gls{ADS} \cready{or focus on other domains, e.g., aerial vehicles.}
        
    \textit{Relevant articles:}
        According to the SLR process suggested by Kitchenham~\cite{kitchenham2004procedures} (cf. Figure~\ref{fig:SLR_process}), we write a review protocol for each relevant paper. We summarize the paper and note relevant aspects wrt. our research questions. We present all relevant articles and categorizations online\footnote{\url{https://doi.org/10.5281/zenodo.7923836}}. 
       
\begin{figure}
            \centering
            \includegraphics[width=\linewidth]{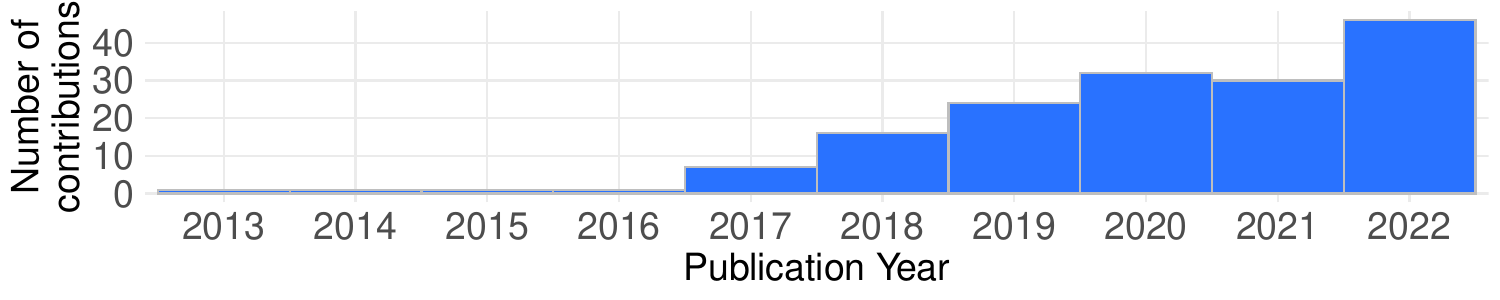}
            \caption{Number of articles that are published per year.}
            \label{fig:years_of_publication}
        \end{figure}

\begin{figure}
            \centering
            \includegraphics[width=\linewidth]{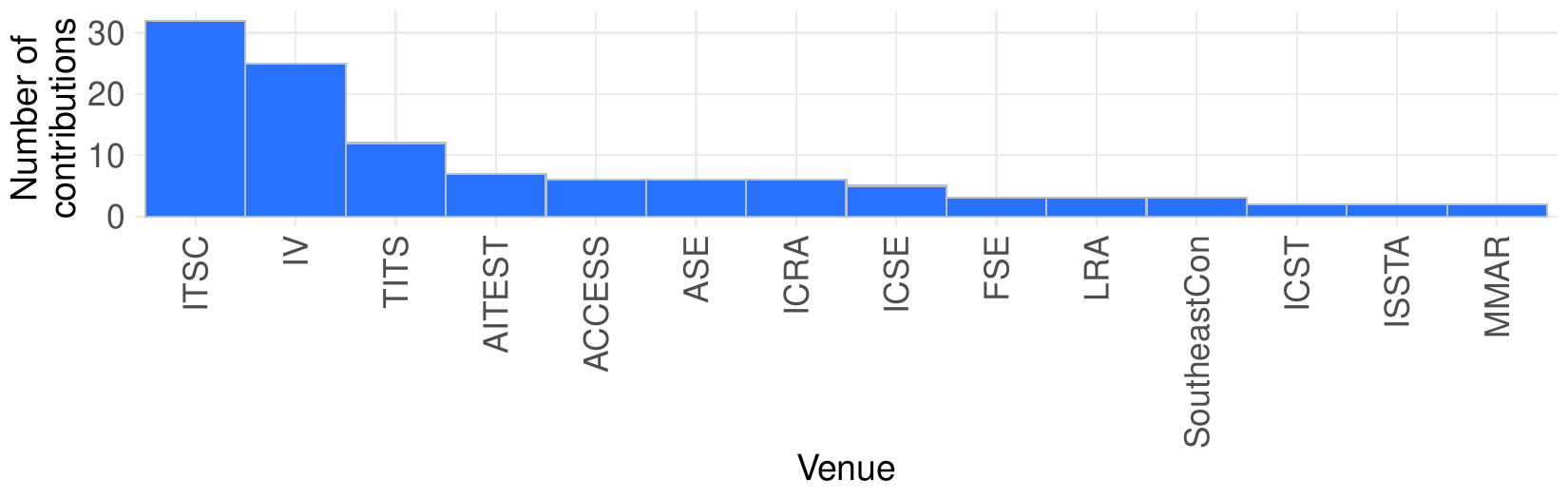}
            \caption{Number of relevant contributions per venue. For the sake of readability, we only present venues that contribute two or more relevant articles.}
            \label{fig:conferences}
        \end{figure}

        Before we analyze relevant articles wrt. our research questions, we examine \cready{their} metadata to make sure that our contribution and research questions are novel and significant.
        First, we consider the publication year of relevant articles. In Figure~\ref{fig:years_of_publication}, we plot a histogram showing the number of published articles per year. Between 2017 and 2022, the number of articles per year increases, while the number of articles per year slightly decreases in 2021.
        However, the increasing trend of published \cready{articles relevant for scenario generation} and a large number of contributions -- even in 2021 -- confirm current research interest in this field.
        Second, we analyze the origin of relevant articles. In Figure~\ref{fig:conferences}, we present a histogram of venues that publish relevant articles; publication years are not considered. Venues that publish relevant articles are in the area of Testing, Software Engineering and (automated) Cyber-physical Systems.
        The main conferences that engage with scenario generation are the \gls{ITSC} and the IEEE \gls{IV}.
        
\section{Discussion} \label{sec:discussion}

    In this section, we discuss and answer our research questions based on our findings.
    We say that an \textit{article} proposes one scenario generation \textit{approach}.
    Regarding RQ~1, we cluster scenario generation \textit{approaches} according to the \textit{technique} that is used. Regarding RQ 2, we focus on the potential and limitations of each \textit{approach}
    to conclude characteristics of scenario generation \textit{techniques}. \\ [1.5ex]
    \noindent\textit{RQ~1: Which techniques are used to generate scenarios for \gls{SBT}?} On a high level, we identify five clusters of scenario generation techniques: (1) data-driven, (2) optimization-based, (3) combinational, (4) expert-based, and (5) random.
        
        \textit{Data-driven} approaches derive scenarios from a database of real-world observation,
        e.g., by reconstruction  \cite{gambi2019generating, montanari2021maneuver}, applying clustering methods \cite{hauer2020clustering}, machine learning \cite{krajewski2019beziervae} or statistical analysis \cite{nalic2019development, de2017assessment}.
        \textit{Optimization-based} scenario generation approaches define a fitness function and generate scenarios to maximize or minimize it. For example, \cite{althoff2018automatic} generates scenarios by minimizing the driveable area of the \gls{SUT}.
        \textit{Combinational} approaches generate scenarios by systematically selecting and combining (atomic) scenario elements \cite{birkemeyer2022feature, rocklage2017automated}.
        \textit{Expert-based} scenario generation uses expert knowledge to parameterize a scenario, e.g., by applying traffic rules \cite{paranjape2020modular} and considering human behavior \cite{yang2020multi}.
        In \textit{random} scenario generation, scenarios are randomly generated, e.g., by randomly selecting parameters \cite{wang2020behavioral} or road elements \cite{medrano2019abstract}.

\begin{figure}
           	\centering
            \includegraphics[width=\linewidth]{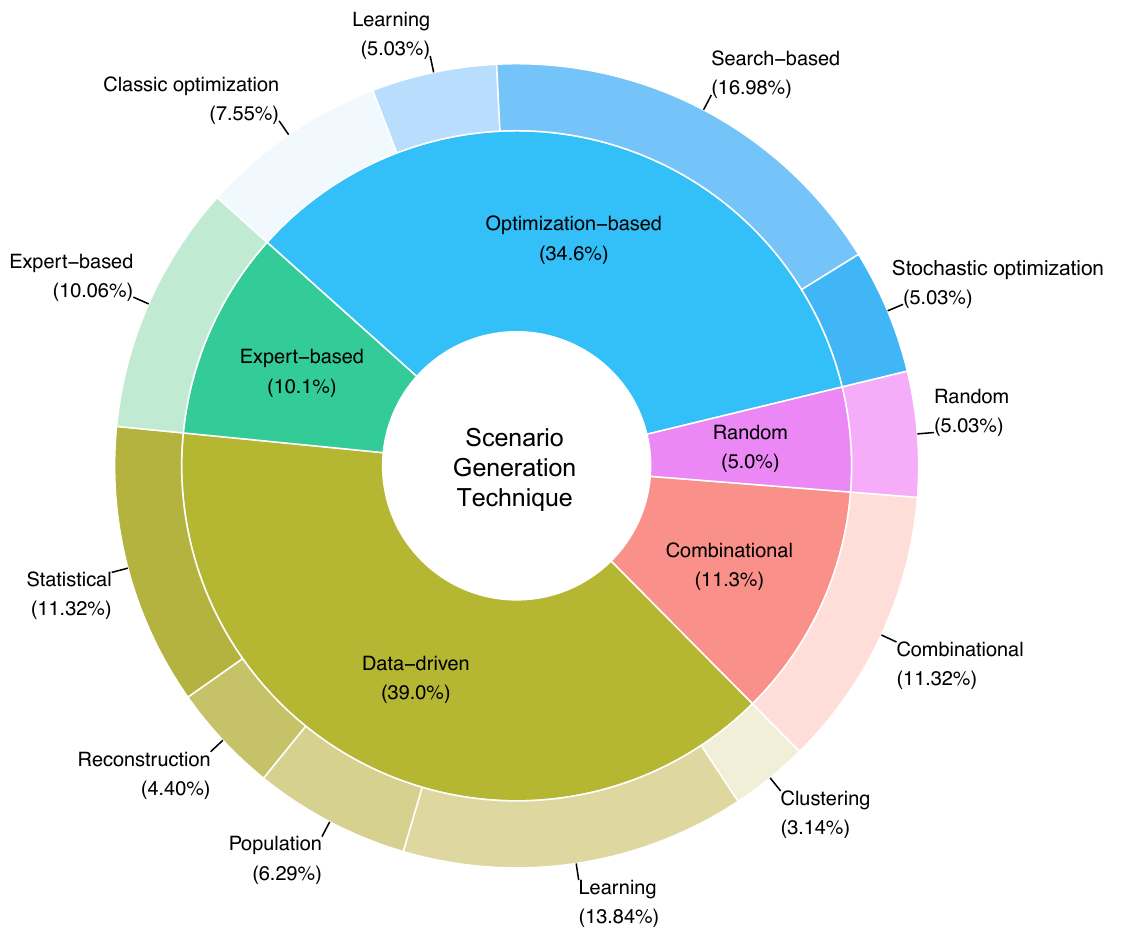}
            \caption{\cready{E}xisting scenario generation techniques (inner circle) and subdivisions (outer circle), in current literature.
            }
            \label{fig:categories}
        \end{figure}

        We assign each scenario generation approach to exactly one technique.
        However, roughly 10\% of scenario generation approaches combine multiple techniques. We assign them to the main technique
        by identifying their main technique by identifying the technique that mostly impacts the parameter selection while transforming logic into concrete scenarios. We classify approaches that contribute simulation tools to support scenario generation by \gls{CAD} as expert-based scenario generation.
        In Figure \ref{fig:categories}, we present the distribution of scenario generation techniques used in the relevant articles in this \gls{SLR}. It is worth noting, that data-driven approaches are the dominating, scenario generation technique. 
        The share of random scenario generation is negligibly small, meaning that scenario generation is predominantly systematic.
        Except for expert-based scenario generation, current approaches generate scenarios automatically so that they have the potential to handle an extremely large space of possible scenarios.        
    \\ [1.5ex]
    \noindent\textit{RQ~2: Do existing scenario generation techniques generate scenarios that fulfill the requirements of the SOTIF-standard?} To ensure reliable statements we explicitly avoid statements based on single articles or small clusters. Hence, regarding RQ~2, we do not consider random scenario generation.
        
\begin{figure}
            \centering
            \subfloat[all approaches]{\includegraphics[trim=0 1cm 25cm 0, clip,height=2.3cm]{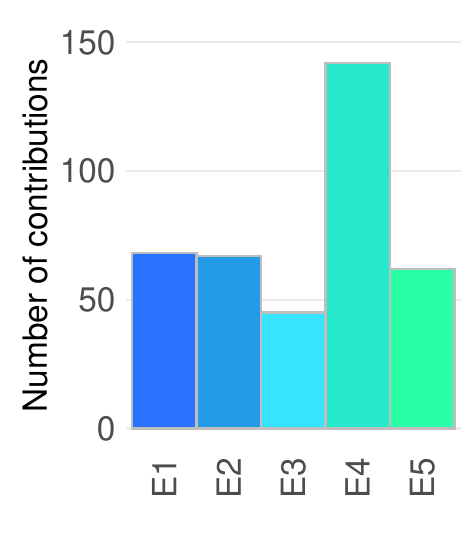}}\hspace{-0.8cm}
            \qquad
            \subfloat[scenario generation techniques]{\includegraphics[trim=8cm 1cm 0 0, clip,height=2.3cm]{pics/bar_level_covered_sum.pdf}}
            \vspace{-0.3\baselineskip}
            \caption{Covered scenarios levels (E1--E5): (a) for all scenario generation approaches and (b) clustered by scenario generation techniques.}
            \label{fig:scenario_levels}
        \end{figure}

        \textit{RQ~2.1: Which details of the real-world are covered?} Although \gls{SOTIF} requires scenarios that represent the real-world or a delimited \gls{ODD} (cf. \textit{Req-1}), \gls{SOTIF} does not provide a metric to evaluate the completeness of scenario coverage.
        To evaluate coverage, we use the scenario level structure by Bagschik~et~al.~\cite{bagschik2018wissensbasierte}. If all possible elements of all scenario levels are covered by generated scenarios, we consider the scenario generation approach to be complete in the sense of modeling the real-world.
        We also determine the number of approaches that contribute to a specific scenario level. We count a contribution if one or more elements of the scenario level are generated and evaluated (e.g., in experiments) or their generation is conceptually discussed. Moreover, we investigate whether a scenario generation approach has the potential to contribute to scenario levels that are not explicitly discussed in the article. Multiple scenario levels per scenario generation approach are possible. In Figure~\ref{fig:scenario_levels}, we present the scenario levels to which scenario generation approaches and techniques contribute. We consider the union of all scenario generation approaches (left) and clustered by scenario generation technique (right).
        The results show that the focus is on scenario level E4; especially, data-driven and optimization-based techniques focus on E4, while combinational and expert-based scenario generation cover all scenario levels approximately equally. On average, scenario level E3 is less covered.
        The focus of current scenario generation approaches on scenario level E4 might be motivated by the fundamental task of human drivers that contains significant challenges: the interaction with other road users on the path control task \cite{donges2015fahrerverhaltensmodelle}. While elements of scenario levels E1 and E2 change slowly and might be part of high definition maps that are deposited in the \gls{ADAS}/\gls{ADS}, elements of scenario level E4 change rapidly and are hard to predict, e.g., due to other traffic users. Thus, developing \gls{ADAS}/\gls{ADS} that interact with level E4 elements are most challenging and require increased \gls{VaV} effort, as reflected by current scenario generation approaches.

        \begin{figure*}[h!]
            \centering
            \includegraphics[width=\textwidth]{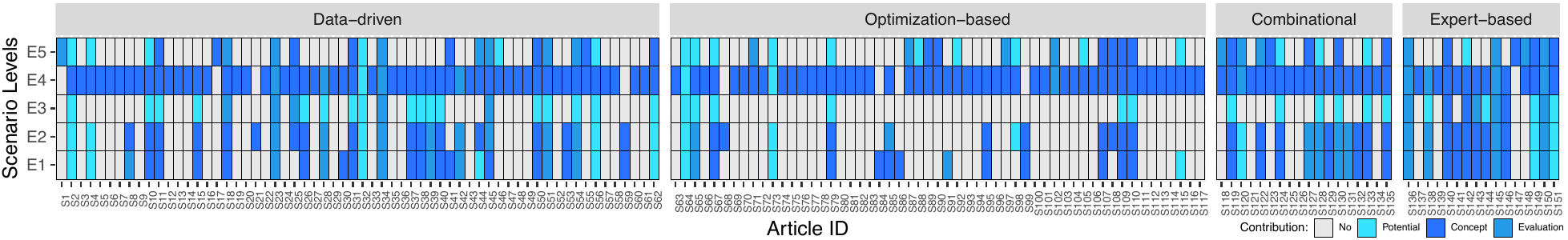}
            \caption{Scenario levels covered per article for articles S1 - S151; separated by scenario generation technique. The color intensity indicates the focus of an article. The more blue boxes are in a vertical line, the more complete a scenario is covered. The more gray boxes are in a vertical line, the more focused is the article on specific scenario elements. 
            }
            \label{fig:levels_per_article}
        \end{figure*}

\begin{figure}
            \centering
            \includegraphics[width=\linewidth]{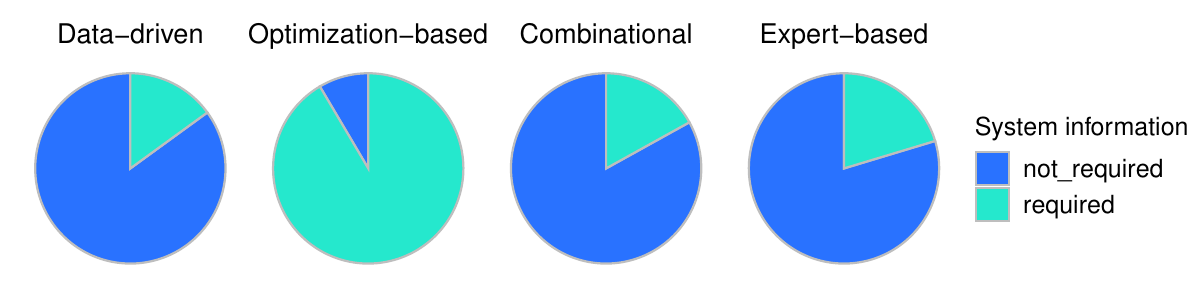}
            \caption{Share of articles that require system information of the \gls{SUT} separated by used scenario generation technique.}
            \label{fig:SUT_required}
        \end{figure}

        In RQ~2.1, we also analyze the completeness of \textit{individual} scenario generation approaches wrt. the real-world: (1) we analyze, whether all scenario levels are covered; approaches that cover all scenario levels have the potential to be real-world complete; thus, (2) we analyze, whether each scenario level is completely considered.
        In Figure \ref{fig:levels_per_article}, we show a heatmap that indicates whether a scenario level is addressed in an article. We distinguish between addressing scenario levels with (a) evaluation, (b) concept or (c) potential. Optimization-based and partially also data-driven techniques only focus on E4, while combinational and expert-based scenario generation address all scenario levels.
        Covering a complete scenario level implies covering a large space of potential elements which is infeasible to assess due to a large number of options.
        In contrast, it is possible to show incompleteness by finding \textit{missing elements}. We find missing elements for each scenario generation approach that address all five scenario levels.
        None of the automated approaches (i.e. data-driven, optimization-based, combinational) covers all elements that are probably relevant for a real-world ODD, such as \textit{gravel}, \textit{yellow lane markings}, \textit{defect bulb in a traffic light}, \textit{motorbikes}, or \textit{snow}, etc. Since we determined these missing example elements with expert knowledge, they can in principle be covered in expert-based scenario generation techniques. However, expert-based scenario generation is limited to human knowledge and does not scale for large-scale scenario generation. We argue that limitations of simulation tools might also lead to missing elements of current scenario generation approaches.
        Regarding \textit{Req-1}, no existing automated scenario generation approach is complete in modeling the real-world and thus, not able to generate scenarios that completely cover it. Covering a delimited \gls{ODD} is, however, possible.  \\ [1.5ex]
    \noindent\textit{RQ~2.2: Are generated scenarios \gls{SUT}-specific?}
        \gls{SOTIF} requires scenarios to minimize the set of (unknown) hazardous scenarios (cf. Req-2). Since this classification is based on both scenario and \gls{SUT}, we determine, whether scenarios explicitly trigger hazardous behavior of the \glspl{SUT} by analyzing whether scenario generation requires information of the \gls{SUT}. As \gls{SUT}, we consider an overall \gls{ADAS} or \pgls{ADS}. 
        Figure \ref{fig:SUT_required} presents the share of scenario generation approaches that do (green) or do not (blue) require system information, separated by scenario generation technique.
        It is remarkable, that optimization-based approaches significantly require system information of the \gls{SUT}, while data-driven, combinational, and expert-based approaches do not.
        In the sense of triggering hazardous behavior, approaches that generate \gls{SUT}-specific scenarios (i.e. wrt. to a specific (type of) \gls{SUT}), have the potential to outperform approaches that generically generate scenarios. Since \gls{SUT}-specific scenarios focus on the \gls{SUT}, they do not holistically represent a scenario space.
        \gls{SUT}-specific and generic scenario generation relate to \textit{exploration} vs. \textit{exploitation} of the overall scenario space. Scenario generation that focuses on \textit{exploration} is associated with generic scenario generation discovering \textit{new} (e.g., unconsidered / unknown) areas of scenario spaces. In contrast, \textit{exploitation} is associated with \gls{SUT}-specific scenario generation, aiming for the \textit{best} (e.g. the most critical / hazardous) scenario. However, exploitation might lead to \textit{local} minima/maxima. Hence, minimizing the area of (unknown) hazardous scenarios (cf. \textit{Req-2}) means carefully balancing exploration vs. exploitation.
        
    \noindent\textit{RQ~2.3: Are scenarios designed to trigger hazardous behavior?}
        As another aspect regarding \textit{Req-2}, we focus on the \gls{SOTIF} classification of generated scenarios. \gls{SOTIF} requires minimizing the set of (unknown) hazardous scenarios; thus, scenario generation approaches that generate hazardous scenarios are highly relevant. We determine which \gls{SOTIF} scenario classes are generated by existing scenario generation approaches; we consider fully automated driving as \gls{SUT}.        
        Due to a missing quantifiable definition of hazardous scenarios, we separate \textit{hazardous} and \textit{not hazardous} scenarios with expert knowledge. We consider a scenario as \textit{hazardous} when it explicitly implements potential triggering conditions such as near-miss accidents; otherwise as \textit{not hazardous}.
        Likewise, an objective definition to distinguish between \textit{known} and \textit{unknown} is missing and depends on the stakeholders' perspective. Each stakeholder (e.g. manufacturer, testing instance, etc.) might have their own database of \textit{known} scenarios.
        In this study, we consider a scenario as \textit{known} if it is based on real-world observations, part of existing databases or designed by expert knowledge; otherwise, a scenario is \textit{unknown}. Since the behavior of \glspl{SUT} is unknown for any kind of \textit{unknown} scenario, we consider \textit{unknown} scenarios as both \textit{hazardous} and \textit{not hazardous} as long as the scenarios do not explicitly implement potential triggering conditions such as near misses.
         \begin{figure*}
            \centering
            \subfloat[all approaches]{\includegraphics[trim=0 0.1cm 34cm 0.2cm, clip,height=4.0cm]{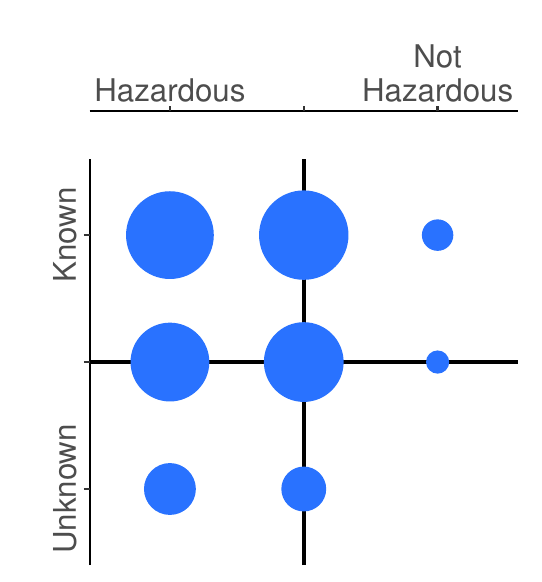}}\hspace{-0.8cm}
            \qquad
            \subfloat[clustered by scenario generation techniques]{\includegraphics[trim=9cm 0.1cm 0 0.2cm, clip,height=4.0cm]{pics/SOTIF_classification_multiple.pdf}}
            \vspace{-0.3\baselineskip}
            \caption{Scenario classification according to the \gls{SOTIF}-standard. We present the four scenario classes that are defined in \gls{SOTIF}. The size and position of the dots indicate the number of approaches that contribute to a specific scenario class.}
            \label{fig:SOTIF_calssification}
            \vspace{-0.2cm}
        \end{figure*}
        In Figure \ref{fig:SOTIF_calssification}, we present the \gls{SOTIF} scenario classes that are generated by all existing scenario generation approaches (left) and separated by scenario generation techniques (right). The blue dots indicate, to which \gls{SOTIF} classes existing scenario generation approaches and techniques contribute.
        Each \gls{SOTIF} class is covered by at least one scenario generation approach.
        Since optimization-based techniques exploit scenario spaces wrt. hazardous scenarios, this technique predominantly leads to hazardous scenarios. 
        Although the logical scenario (at the beginning of the optimization) is known, after the optimization process, the concrete scenario can be both known and unknown.
        Combinational techniques explore the overall scenario space. Depending on the strategy for combining atomic scenario elements, resulting scenarios are both known/unknown as well as hazardous/not hazardous.
        Data-driven techniques explore the scenario space by collecting real-world data, e.g., by sensor-equipped vehicles. Reconstruction leads to known scenarios; unknown scenarios are possible by varying parameters. Data-driven techniques generate realistic scenarios, while optimization-based and combinational techniques generate artificial scenarios.
        It needs to be validated whether artificial scenarios properly represent the real-world. Expert-based techniques focus on known/hazardous scenarios.

\section{Threats to Validity}
    Although \pgls{SLR} is a sound and established method to determine the state-of-the-art, there are threats to its validity.

    \textit{Internal Validity:}
        A threat to internal validity is that we misunderstand articles so that we incorrectly classify them wrt. their scenario generation technique, covered scenario levels, required system information, or \gls{SOTIF} scenario classes. To mitigate this threat, we do not make conclusions based on single or small clusters of articles. Moreover, we double-check article classification by two independent researchers and make it traceable by sharing details online.
        \cready{Another threat is that we collected articles from unsuitable sources. To mitigate this threat, we focus on databases and search engines that are commonly used within technical scientific research.}
        \cready{A last aspect that threatens the correctness of our results is that we defined unsuitable search terms to collect literature from databases. To mitigate this threat, we faithfully and iteratively specified the area of interest with expert knowledge from two independent researchers. We determine synonyms and cover arbitrary grammatical forms.}

    \cready{\textit{External Validity:}}
        A threat to the \cready{external} validity is that the requirements of \gls{SOTIF} are an unsuitable scope to discuss existing scenario generation \cready{and thus, the results might not be generalizable for \gls{VaV} of \gls{ADAS}/\gls{ADS}}.
        \cready{However}, we argue that \gls{SOTIF} is an official standard for \gls{VaV} of \gls{ADAS}/\gls{ADS} and is widely accepted in the automotive domain.
        We faithfully derived requirements from the latest \gls{SOTIF} version.


\section{Conclusion \& Future Research} \label{sec:conclusion}
    The \gls{SOTIF}-standard introduces \gls{SBT} to verify and validate \gls{ADAS}/\gls{ADS}. However, \gls{SOTIF} does not define methods to generate or assess scenarios which hinders the practical applicability of \gls{SOTIF}. In this article, we analyze existing \cready{approaches to generate concrete scenarios} and evaluate whether generated scenarios comply with the requirements implied by the \gls{SOTIF}-standard. We derived five clusters of scenario generation techniques: \textit{data-driven}, \textit{optimization-based}, \textit{combinational}, \textit{expert-based}, and \textit{random}.

    \textit{Data-driven} techniques generate generic and realistic scenarios based on collected data.
    \textit{Optimization-based} techniques automatically exploit scenario spaces wrt. scenarios that trigger hazardous behavior of the \gls{SUT}.
    In contrast to exploiting a scenario space, \textit{combinational} scenario generation automatically explores the overall scenario space by combining atomic scenario elements.
    \textit{Expert-based} techniques are not automatic and do not scale for real-world problems.
    \textit{Random} scenario generation is an automatic, but unsystematic approach which is only marginal in the literature.
    
    Regarding the \gls{SOTIF}-standard, we need to generate scenarios that cover \pgls{ODD} and trigger hazardous behavior of the \gls{SUT}. Existing scenario generation approaches, however, do not fulfill both requirements respectively. 
    First, since complete coverage of the real-world is an infeasible object to reach, we might counteract coverage completeness (cf. \textit{Req-1}) by scenarios that do not cover, but represent an \gls{ODD} or the real-world.
    Combinational techniques have the potential to build representative subsets and data-driven scenario generation stands out in generating realistic scenarios.
    Second, scenarios need to trigger hazardous behavior of the \gls{SUT} (cf. \textit{Req-2}), which is the focus of optimization-based scenario generation.
    \cready{However, optimization techniques focus on (local) minima/maxima and do not explore a scenario space; while combinational scenario generation does not explicitly select scenarios that trigger hazardous behavior and data-driven approaches heavily rely on the input-dataset.}
    Hence, as a direction for future work, we strongly suggest combining existing scenario generation techniques. We need to generate scenario suites that carefully balance representativeness (\textit{combinational}), hazardous triggering conditions (\textit{optimization-based}), and realism (\textit{data-driven}). \cready{Birkemeyer et al. \cite{birkemeyer2023semi}, for example, suggests a concept to integrate optimization techniques in combinational scenario generation and indicate the potential of combining both techniques.}
    
    The set of possible scenarios is extremely large and thus, the set of scenarios that represent \pgls{ODD} or the real-world. Hence, we need to consider scalability during the scenario generation process. Since the scenario generation approaches data-driven, optimization-based and combinational automatically generate scenarios, these techniques have the potential to practically generated scenarios for \gls{VaV} of \gls{ADAS}/\gls{ADS}.
    Finally, to balance the potential of multiple scenario generation techniques as suggested before, we need to define metrics that assess scenario suites, independently from the scenario generation process. In the sense of \gls{VaV}, we could apply mutation testing to assess the ability of a scenario suite to detect errors in the \gls{SUT}~\cite{birkemeyer2022feature}.
    
    To sum up, scenarios generated with the existing scenario generation approaches do not comply with requirements implied by the \gls{SOTIF}-standard. To close this gap, we propose the following directions for future research:

    \begin{itemize}
        \item Combining existing scenario generation techniques has potential to generate scenarios that fulfill \gls{SOTIF} requirements.
        \item Scenario generation needs to be scalable to generate scenarios suites that represent large-scale \glspl{ODD} or the real-world.
        \item A metric that assess scenario suites independently from the scenario generation process is required.
    \end{itemize}
    
    \section*{Acknowledgments}
        This work has been partially funded by the Ph.D. program "Responsible AI in the Digital Society" funded by the Ministry for Science and Culture of Lower Saxony.
        This work has been partially supported by the research project SofDCar (19S21002), which is funded by the German Federal Ministry for Economic Affairs and Climate Action.




\bibliographystyle{IEEEtran}
\balance
\bibliography{bib.bib}

\begin{thebibliography}{10}
\providecommand{\url}[1]{#1}
\csname url@samestyle\endcsname
\providecommand{\newblock}{\relax}
\providecommand{\bibinfo}[2]{#2}
\providecommand{\BIBentrySTDinterwordspacing}{\spaceskip=0pt\relax}
\providecommand{\BIBentryALTinterwordstretchfactor}{4}
\providecommand{\BIBentryALTinterwordspacing}{\spaceskip=\fontdimen2\font plus
\BIBentryALTinterwordstretchfactor\fontdimen3\font minus
  \fontdimen4\font\relax}
\providecommand{\BIBforeignlanguage}[2]{{%
\expandafter\ifx\csname l@#1\endcsname\relax
\typeout{** WARNING: IEEEtran.bst: No hyphenation pattern has been}%
\typeout{** loaded for the language `#1'. Using the pattern for}%
\typeout{** the default language instead.}%
\else
\language=\csname l@#1\endcsname
\fi
#2}}
\providecommand{\BIBdecl}{\relax}
\BIBdecl

\bibitem{iso26262}
\emph{Road vehicles — Functional safety}, Std. ISO 26\,262:2018, 2018.

\bibitem{iso21448}
\emph{Road vehicles — Safety of the intended functionality}, Std. ISO
  21\,448:2022, 2022.

\bibitem{uneceR157}
``{UN} regulation no 157 – uniform provisions concerning the approval of
  vehicles with regards to automated lane keeping systems [2021/389],''
  \url{http://data.europa.eu/eli/reg/2021/389/oj}, pp. 75--137, Mar 2021.

\bibitem{ulbrich2015defining}
S.~Ulbrich, T.~Menzel, A.~Reschka, F.~Schuldt, and M.~Maurer, ``Defining and
  substantiating the terms scene, situation, and scenario for automated
  driving,'' in \emph{IEEE International Conference on Intelligent
  Transportation Systems (ITSC)}, 2015, pp. 982--988.

\bibitem{Bagschik2018a}
G.~Bagschik, T.~Menzel, C.~K{\"o}rner, and M.~Maurer, ``Wissensbasierte
  szenariengenerierung f{\"u}r betriebsszenarien auf deutschen autobahnen,'' in
  \emph{Workshop Fahrerassistenzsysteme und automatisiertes Fahren. Bd},
  vol.~12, 2018.

\bibitem{annexC(2022)5402}
``Annexes to the commission implementing regulation - c(2022)5402,''
  \url{https://ec.europa.eu/info/law/better-regulation/have-your-say/initiatives/12152-Automated-cars-technical-specifications_en},
  2022.

\bibitem{iso21448pas}
\emph{Road vehicles — Safety of the intended functionality}, Std. ISO/PAS
  21\,448:2019, 2019.

\bibitem{birkemeyer2022fahren}
L.~Birkemeyer, M.~Delventhal, I.~Schaefer, and F.~Schmieder, ``Wann fahren wir
  autonom? eine untersuchung aus technischer und rechtlicher sicht.'' in
  \emph{Software Engineering 2022 Workshops}.\hskip 1em plus 0.5em minus
  0.4em\relax Gesellschaft f{\"u}r Informatik eV, 2022.

\bibitem{gambi2019generating}
A.~Gambi, T.~Huynh, and G.~Fraser, ``Generating effective test cases for
  self-driving cars from police reports,'' in \emph{Proceedings of the 27th ACM
  Joint Meeting on European Software Engineering Conference and Symposium on
  the Foundations of Software Engineering}, 2019, pp. 257--267.

\bibitem{montanari2021maneuver}
F.~Montanari, C.~Stadler, J.~Sichermann, R.~German, and A.~Djanatliev,
  ``Maneuver-based resimulation of driving scenarios based on real driving
  data,'' in \emph{IEEE Intelligent Vehicles Symposium (IV)}, 2021, pp.
  1124--1131.

\bibitem{hauer2020clustering}
F.~Hauer, I.~Gerostathopoulos, T.~Schmidt, and A.~Pretschner, ``Clustering
  traffic scenarios using mental models as little as possible,'' in \emph{IEEE
  Intelligent Vehicles Symposium (IV)}, 2020, pp. 1007--1012.

\bibitem{krajewski2019beziervae}
R.~Krajewski, T.~Moers, A.~Meister, and L.~Eckstein, ``B{\'e}ziervae: Improved
  trajectory modeling using variational autoencoders for the safety validation
  of highly automated vehicles,'' in \emph{IEEE Intelligent Transportation
  Systems Conference (ITSC)}, 2019, pp. 3788--3795.

\bibitem{nalic2019development}
D.~Nalic, A.~Eichberger, G.~Hanzl, M.~Fellendorf, and B.~Rogic, ``Development
  of a co-simulation framework for systematic generation of scenarios for
  testing and validation of automated driving systems,'' in \emph{IEEE
  Intelligent Transportation Systems Conference (ITSC)}, 2019, pp. 1895--1901.

\bibitem{de2017assessment}
E.~de~Gelder and J.-P. Paardekooper, ``Assessment of automated driving systems
  using real-life scenarios,'' in \emph{IEEE Intelligent Vehicles Symposium
  (IV)}, 2017, pp. 589--594.

\bibitem{althoff2018automatic}
M.~Althoff and S.~Lutz, ``Automatic generation of safety-critical test
  scenarios for collision avoidance of road vehicles,'' in \emph{IEEE
  Intelligent Vehicles Symposium (IV)}, 2018, pp. 1326--1333.

\bibitem{birkemeyer2022feature}
L.~Birkemeyer, T.~Pett, A.~Vogelsang, C.~Seidl, and I.~Schaefer,
  ``Feature-interaction sampling for scenario-based testing of advanced driver
  assistance systems,'' in \emph{Proceedings of the 16th International Working
  Conference on Variability Modelling of Software-Intensive Systems (VaMoS)},
  2022, pp. 1--10.

\bibitem{rocklage2017automated}
E.~Rocklage, H.~Kraft, A.~Karatas, and J.~Seewig, ``Automated scenario
  generation for regression testing of autonomous vehicles,'' in \emph{IEEE
  International Conference on Intelligent Transportation Systems (ITSC)}, 2017,
  pp. 476--483.

\bibitem{paranjape2020modular}
I.~Paranjape, A.~Jawad, Y.~Xu, A.~Song, and J.~Whitehead, ``A modular
  architecture for procedural generation of towns, intersections and scenarios
  for testing autonomous vehicles,'' in \emph{IEEE Intelligent Vehicles
  Symposium (IV)}, 2020, pp. 162--168.

\bibitem{yang2020multi}
D.~Yang, K.~Redmill, and {\"U}.~{\"O}zg{\"u}ner, ``A multi-state social force
  based framework for vehicle-pedestrian interaction in uncontrolled pedestrian
  crossing scenarios,'' in \emph{IEEE Intelligent Vehicles Symposium (IV)},
  2020, pp. 1807--1812.

\bibitem{wang2020behavioral}
X.~Wang, Y.~Dong, S.~Xu, H.~Peng, F.~Wang, and Z.~Liu, ``Behavioral competence
  tests for highly automated vehicles,'' in \emph{IEEE Intelligent Vehicles
  Symposium (IV)}, 2020, pp. 1323--1329.

\bibitem{medrano2019abstract}
C.~Medrano-Berumen and M.~I. Akba{\c{s}}, ``Abstract simulation scenario
  generation for autonomous vehicle verification,'' in
  \emph{SoutheastCon}.\hskip 1em plus 0.5em minus 0.4em\relax IEEE, 2019, pp.
  1--6.

\bibitem{menzel2019functional}
T.~Menzel, G.~Bagschik, L.~Isensee, A.~Schomburg, and M.~Maurer, ``From
  functional to logical scenarios: Detailing a keyword-based scenario
  description for execution in a simulation environment,'' in \emph{IEEE
  Intelligent Vehicles Symposium (IV)}, 2019, pp. 2383--2390.

\bibitem{stellet2015testing}
J.~E. Stellet, M.~R. Zofka, J.~Schumacher, T.~Schamm, F.~Niewels, and J.~M.
  Z{\"o}llner, ``Testing of advanced driver assistance towards automated
  driving: A survey and taxonomy on existing approaches and open questions,''
  in \emph{IEEE International Conference on Intelligent Transportation Systems
  (ITSC)}, 2015, pp. 1455--1462.

\bibitem{nalic2020scenario}
D.~Nalic, T.~Mihalj, M.~B{\"a}umler, M.~Lehmann, A.~Eichberger, and
  S.~Bernsteiner, ``Scenario based testing of automated driving systems: A
  literature survey,'' in \emph{FISITA web Congress}, vol.~10, 2020.

\bibitem{gangopadhyay2019identification}
B.~Gangopadhyay, S.~Khastgir, S.~Dey, P.~Dasgupta, G.~Montana, and P.~Jennings,
  ``Identification of test cases for automated driving systems using bayesian
  optimization,'' in \emph{IEEE Intelligent Transportation Systems Conference
  (ITSC)}, 2019, pp. 1961--1967.

\bibitem{ding2023survey}
W.~Ding, C.~Xu, M.~Arief, H.~Lin, B.~Li, and D.~Zhao, ``A survey on
  safety-critical driving scenario generation—a methodological perspective,''
  \emph{IEEE Transactions on Intelligent Transportation Systems}, 2023.

\bibitem{ma2022verification}
Y.~Ma, C.~Sun, J.~Chen, D.~Cao, and L.~Xiong, ``Verification and validation
  methods for decision-making and planning of automated vehicles: A review,''
  \emph{IEEE Transactions on Intelligent Vehicles}, 2022.

\bibitem{zhang2022finding}
X.~Zhang, J.~Tao, K.~Tan, M.~Torngren, J.~M.~G. Sanchez, M.~R. Ramli, X.~Tao,
  M.~Gyllenhammar, F.~Wotawa, N.~Mohan \emph{et~al.}, ``Finding critical
  scenarios for automated driving systems: A systematic mapping study,''
  \emph{IEEE Transactions on Software Engineering}, 2022.

\bibitem{zhong2021survey}
Z.~Zhong, Y.~Tang, Y.~Zhou, V.~d.~O. Neves, Y.~Liu, and B.~Ray, ``A survey on
  scenario-based testing for automated driving systems in high-fidelity
  simulation,'' \emph{arXiv preprint arXiv:2112.00964}, 2021.

\bibitem{nascimento2019systematic}
A.~M. Nascimento, L.~F. Vismari, C.~B. S.~T. Molina, P.~S. Cugnasca, J.~B.
  Camargo, J.~R. de~Almeida, R.~Inam, E.~Fersman, M.~V. Marquezini, and A.~Y.
  Hata, ``A systematic literature review about the impact of artificial
  intelligence on autonomous vehicle safety,'' \emph{IEEE Transactions on
  Intelligent Transportation Systems}, vol.~21, no.~12, pp. 4928--4946, 2019.

\bibitem{alawadhi2020systematic}
M.~Alawadhi, J.~Almazrouie, M.~Kamil, and K.~A. Khalil, ``A systematic
  literature review of the factors influencing the adoption of autonomous
  driving,'' \emph{International Journal of System Assurance Engineering and
  Management}, vol.~11, no.~6, pp. 1065--1082, 2020.

\bibitem{jing2020agent}
P.~Jing, H.~Hu, F.~Zhan, Y.~Chen, and Y.~Shi, ``Agent-based simulation of
  autonomous vehicles: A systematic literature review,'' \emph{IEEE Access},
  vol.~8, pp. 79\,089--79\,103, 2020.

\bibitem{tahir2020coverage}
Z.~Tahir and R.~Alexander, ``Coverage based testing for {V\&V} and safety
  assurance of self-driving autonomous vehicles: A systematic literature
  review,'' in \emph{IEEE International Conference On Artificial Intelligence
  Testing (AITest)}, 2020, pp. 23--30.

\bibitem{karunakaran2022challenges}
D.~Karunakaran, J.~S. Berrio, S.~Worrall, and E.~Nebot, ``Challenges of testing
  highly automated vehicles: A literature review,'' in \emph{IEEE International
  Conference on Recent Advances in Systems Science and Engineering (RASSE)},
  2022, pp. 1--8.

\bibitem{schutt20231001}
B.~Sch{\"u}tt, J.~Ransiek, T.~Braun, and E.~Sax, ``1001 ways of scenario
  generation for testing of self-driving cars: A survey,'' \emph{arXiv preprint
  arXiv:2304.10850}, 2023.

\bibitem{kitchenham2004procedures}
B.~Kitchenham, ``Procedures for performing systematic reviews,'' \emph{Keele,
  UK, Keele University}, vol.~33, no. 2004, pp. 1--26, 2004.

\bibitem{bagschik2018wissensbasierte}
G.~Bagschik, T.~Menzel, C.~K{\"o}rner, and M.~Maurer, ``Wissensbasierte
  {S}zenariengenerierung f{\"u}r {B}etriebsszenarien auf deutschen
  {A}utobahnen,'' in \emph{Workshop Fahrerassistenzsysteme und automatisiertes
  Fahren. Bd}, vol.~12, 2018, p.~12.

\bibitem{donges2015fahrerverhaltensmodelle}
E.~Donges, ``Fahrerverhaltensmodelle,'' in \emph{Handbuch
  Fahrerassistenzsysteme}.\hskip 1em plus 0.5em minus 0.4em\relax Springer,
  2015, pp. 17--26.

\bibitem{birkemeyer2023semi}
L.~Birkemeyer, J.~Fuchs, A.~Gambi, and I.~Schaefer, ``{SOTIF}-compliant
  scenario generation using semi-concrete scenarios and parameter sampling,''
  in \emph{IEEE International Conference on Intelligent Transportation Systems
  (ITSC)}, 2023.

\end{thebibliography}

\end{document}